\documentclass[sigconf]{acmart}
\renewcommand\footnotetextcopyrightpermission[1]{} 
\AtBeginDocument{%
  \providecommand\BibTeX{{%
    \normalfont B\kern-0.5em{\scshape i\kern-0.25em b}\kern-0.8em\TeX}}}

\setcopyright{none}
\copyrightyear{2019}
\acmYear{2019}

\acmConference[Preprint, July 2019]{Preprint, July 2019}

\begin{document}

\sloppy

\title{Tracking sex: The implications of widespread sexual data leakage and tracking on porn websites (Preprint, July 2019)}


\author{Elena Maris}
\affiliation{Microsoft Research}
\email{elena.maris@microsoft.com}

\author{Timothy Libert}
\affiliation{Carnegie Mellon University}
\email{timlibert@cmu.edu}

\author{Jennifer Henrichsen}
\affiliation{University of Pennsylvania}
\email{jennifer.henrichsen@asc.upenn.edu}

\renewcommand{\shortauthors}{Maris, Libert, and Henrichsen}

\begin{abstract}
This paper explores tracking and privacy risks on pornography websites. Our analysis of 22,484 pornography websites indicated that 93\% leak user data to a third party. Tracking on these sites is highly concentrated by a handful of major companies, which we identify. We successfully extracted privacy policies for 3,856 sites, 17\% of the total. The policies were written such that one might need a two-year college education to understand them. Our content analysis of the sample's domains indicated 44.97\% of them expose or suggest a specific gender/sexual identity or interest likely to be linked to the user. We identify three core implications of the quantitative results: 1) the unique/elevated risks of porn data leakage versus other types of data, 2) the particular risks/impact for vulnerable populations, and 3) the complications of providing consent for porn site users and the need for affirmative consent in these online sexual interactions.
\end{abstract}




\maketitle

\section{Introduction}
\begin{quote}
One evening, `Jack' decides to view porn on his laptop. He enables `incognito' mode in his browser, assuming his actions are now private\footnote{Private browsing is used more often when viewing `adult' content; however, users `overestimate the protection from online tracking and targeted advertising,' which is scant (Habib et al., 2018: 159).}. He pulls up a site and scrolls past a small link to a privacy policy. Assuming a site with a privacy policy will protect his personal information\footnote{See Turow et al. (2015a).}, Jack clicks on a video. What Jack does not know is that incognito mode only ensures his browsing history is not stored on his computer. The sites he visits, as well as any third-party trackers, may observe and record his online actions. These third-parties may even infer Jack's sexual interests from the URLs of the sites he accesses. They might also use what they have decided about these interests for marketing or building a consumer profile. They may even sell the data. Jack has no idea these third-party data transfers are occurring as he browses videos. His assumption that porn websites will protect his information, along with the reassurance of the `incognito' mode icon on his screen, provide Jack a fundamentally misleading sense of privacy as he consumes porn online.
\end{quote}

The above hypothetical scenario occurs frequently in reality and is indicative of the widespread data leakage and tracking that can occur on porn sites. In 2017, Pornhub, one of the largest porn websites\footnote{Pornhub  is easily one of the largest porn sites in terms of content; in its  first 10 years, more than 10 million videos were uploaded on the site  (Pornhub, 2017).}, received 28.5 billion visits, with users performing 50,000 searches per second on the site (Pornhub, 2018). Statistics vary as to the amount of overall porn activity on the internet, but a 2017 report indicated porn sites get more visitors each month than Netflix, Amazon, and Twitter combined, and that `30\% of all the data transferred across the internet is porn,' with site YouPorn using six times more bandwidth than Hulu (Kleinman, 2017). While there is much scholarly attention on internet use and privacy in general, there has been less research on the specific privacy implications of online porn use. Considering that porn websites are among the most visited on the Internet (Alexa, 2018b), it is imperative to attend to the specific privacy concerns of online porn consumption. Most crucially, porn consumption data is \emph{sexual} data, and thus constitutes an especially sensitive type of online data users likely wish to keep private.

Revelations about such data represent specific threats to personal safety and autonomy in any society that polices gender and sexuality. In this article, we demonstrate through the study of 22,484 pornography websites that people who visit such sites may have their sexual interests inferred by third-parties that surreptitiously track web browsing, often without user notice or consent. We provide quantitative results that reveal extensive privacy issues on pornography websites and highlight three core implications of our findings: 1) the unique and elevated risks of porn data leakage versus other types of data, 2) the targeting of `difference' and the likelihood that the tracking of sexual data will especially impact vulnerable populations, and 3) the complications of giving consent to data collection and tracking for porn site users, and how these problematic understandings of consent mirror more general misconceptions and power imbalances of interpersonal sexual consent.

\section{Related Work}

\subsection{Porn Uses, Identity, and `Sexual Interests'}

Pornography and sexually explicit material related to sex, sexual orientation, gender performance, and sexual interests have long served as sources of information, identity formation, support, and community. This has particularly been true for those whose sexual interests are deemed deviant or abnormal, and thus must be explored privately. Gross (2001: 221) explains, `..sexual images and stories have generally been officially condemned while privately enjoyed. They also have offered channels for the vicarious expression and satisfaction of minority interests that are difficult, embarrassing, and occasionally illegal to indulge in reality\ldots' Porn can provide community for those in areas hostile toward their identity  (Gross, 2001). Despite online porn's affordances for community, it remains tied up in extant issues around power, agency and representation in traditional porn industries (Mowlabocus, 2010).

We center private access to online porn as important to a queer, feminist, sex-positive politics of gender and sexuality, and central to community-building and free and safe sexual expression:

\begin{quote}
The existence of these sexual images is a threat to those who guard the ramparts of the sexual reservation. Visible lesbian or gay (or any unconventional) sexuality undermines the unquestioned normalcy of the status quo and opens up the possibility of making choices that people might never have otherwise considered. (Gross, 2001: 223)
\end{quote}

\noindent When sex acts and identities are labeled abnormal \emph{or} normal,\emph{all} are vulnerable. Sloop (2004: 8) notes `sex positive' means, ``to think of sexual practices and sexuality as being organized into systems of power that must be transgressed if we are to undermine the constraining dimensions of culture on our behavior,'' and, according to Smith and Attwood (2014: 13) is ``\ldots often associated with opposition to the regulation of sexual practices, the censorship of sexual representations and restrictions on sex education.'' Herein, we take such a `sex positive' view of porn and access to online pornography. While acknowledging the many racist, misogynistic, heteronormative and other problematic histories and themes in pornography and its production, distribution and consumption\footnote{See  Williams (2004) and Smith and Attwood (2014), on the prominent  theories, debates and critiques of (online) pornography.}, our work recognizes the ubiquity and permanence of porn and its many uses and social functions, and the danger of societal, state, and institutional narratives that might work to discipline gender and sex.

Researchers are learning the many uses of porn, complicating simplistic notions of what porn is `for.' Porn consumption does not necessarily equate to sexual identity, preference, pleasure, interest, or fetish. For example, one may consume gay porn but not identify as gay. Barker (2014: 149) notes the ``rich variety'' of reported reasons for viewing porn, including: ``for reconnection with my body, to get in the mood with my partner, for recognition of my sexual interests, to see things I might do, to see things I can't do, to see things I wouldn't do, to see things I shouldn't do, for a laugh\ldots,'' and more. Sexual playfulness is an important means of exploring changing pleasures and preferences outside of strict categorizations of identity that can stigmatize some interests (Paasonen, 2018; Tiidenberg and Paasonen, 2018). Thus, when we note a porn site or user's `sexual interest' or `sexual data' is revealed or could be inferred in tracking porn site visits, we do so with the knowledge that porn serves a variety of uses and content consumed does not explicitly indicate a person's sexual or gender identity, interest, desire, or affinity\footnote{It also doesn't necessarily reveal actions  by the assumed device owner; porn consumption can occur on someone's  device without their knowledge.}. Further, the site URLs often suggest specific genders and/or sexual preferences, genres, and acts found in the site \emph{content}. However, we believe if individuals' porn use is involuntarily exposed, such nuanced, sex-positive understandings of porn and sexual interest will likely \emph{not} figure into many outside readings of user activities. Thus, we center the ability to privately consume online porn as a right to \emph{sexual privacy}, which Citron (2019: 1898, 1901) notes, ``is concerned with sexual autonomy,self-determination, and dignity\ldots'' and ``\ldots the extent to which others have access to and information about people's \ldots{} sexual desires, fantasies, and thoughts\ldots''

\subsection{Online Tracking and Privacy}

Although users may perceive a website or app as a single entity (often the address in their browsers), many sites and apps include code from other parties of which users are typically unaware (Libert, 2015). Such ``third-party'' code can allow companies to monitor the actions of users without their knowledge or consent and build detailed profiles of their habits and interests. Such profiles are often used for targeted advertising, for example, by showing ads for dog food to dog owners (Turow, 2012). Many websites and apps have revenue sharing agreements with third-party advertising networks and gain direct monetary benefit from including third-party code (Turow, 2012). However, tracking users on websites \emph{without} advertisements can provide additional insights into their habits, and online advertising companies like Facebook and Google offer web developers a range of ``free'' non-advertising services subsidized by allowing these companies to track users (Libert, 2015). For example, a developer may include the Facebook ``Like'' button on a website to facilitate sharing content, which allows Facebook to track the activities of all visitors - Facebook users or not. Decades of research have demonstrated a variety of types of third-party tracking are endemic on both web and mobile devices (Felten and Schneider, 2000; Krishnamurthy and Wills, 2006; Libert, 2015; Englehardt and Narayanan, 2016; Binns et al., 2018).

The impacts of this tracking extend far beyond selling dog food. A significant body of literature has addressed the social implications of online consumer surveillance, including users' attitudes about being tracked (Barth and de Jong, 2017; Custers et al., 2014), the mechanisms behind data mining and tracking (Kennedy, 2016), how developers define and design for privacy (Greene and Shilton, 2018), and surveillance as a technology of control within capitalism (Campbell and Carlson, 2002). Collecting and tracking data are often framed as ways to `know' quantitatively unknowable and often morally charged constructions like who or what is `average,' `normal,' or `healthy' (Ruckenstein and Pantzer, 2017: 408). Indeed, van Dijck (2014: 198) states that `dataism' demonstrates, ``widespread \emph{belief} in the objective quantification and potential tracking of \ldots human behavior and sociality\ldots{}(and) also involves \emph{trust} in the (institutional) agents that collect, interpret, and share (meta)data\ldots''

Despite the normalization of tracking, survey research consistently demonstrates that users do not enjoy being tracked online (Cranor et al, 2000; Turow et al, 2015a). Nissenbaum (2010: 2) argues, ``What people care most about is not simply \emph{restricting} the flow of information but ensuring that it flows \emph{appropriately}.'' Privacy policies, the primary means for users to learn about tracking, have been consistently found inadequate due to users not understanding their \emph{purpose} (Smith, 2014), difficulty understanding the dense legalese in which they are written (McDonald and Cranor, 2008), and that such policies fail to disclose 85\% of observed instances of third-party tracking (Libert,2018). Despite this, the online advertising industry asserts users can ``opt-out'' of such tracking under a self-regulatory framework referred to as `notice and choice' or `notice and consent' (Baruh and Popescu, 2017). While some point to a `privacy paradox' between users' expressed privacy preferences and their actual behaviors, one compelling explanation is that `notice and consent' is so confusing users are unable to `opt-out' even if they wish to do so (Smith, 2014). It is important to note the new General Data Protection Regulation (GDPR) in the European Union is designed to curb the practices described above by forbidding many forms of third-party tracking without affirmative consent from users (Libert and Nielsen, 2018). However, the GDPR does not apply world-wide and its impacts are not yet clear.

\section{Research Questions}

This paper aims to ascertain the potential for surveillance and tracking of pornography website visitors and their associated sexual data. Further, it explores theoretically-informed implications of data leakage, tracking, and other security concerns related to privacy and online pornography consumption. Although we use a global sample of porn sites (loaded in the U.S.), and will note at points in this article where global contexts might be especially relevant, we approach this project from a U.S.-based culture, policy, and privacy perspective. The research was conducted with the following research questions:

\begin{itemize}

\item \textbf{RQ1:} To what extent do pornography websites potentially reveal user data and allow for third-party identification and tracking?

\item \textbf{RQ2:} What entities/organizations tend to have the most access to this data? Do the sites' privacy policies disclose tracking and the organizations with access to their data?

\item \textbf{RQ3:} What is the potential for pornography website users' sexual interests to be revealed or inferred by such surveillance and tracking?

\item \textbf{RQ4:} What are the potential implications of porn website surveillance and tracking? What consequences for users can be drawn from the results, especially informed by theories of gender, sexuality and privacy, as well as relevant prior cases?

\end{itemize}

\section{Methodology}

\subsection{Sample}

In March 2018, we used a U.S.-based computer to analyze 22,484 pornography websites to identify the third-parties which may be able to infer users' sexual interests, and whether privacy policies provide a sufficient vehicle for obtaining meaningful consent to tracking. To create our population of pornography websites, we downloaded the homepages of the one million most popular websites identified by the Alexa service\footnote{Alexa, a subsidiary of Amazon, provides website  traffic metrics and rankings `based on the browsing behavior of people  in {[}a{]} global data panel which is a sample of all internet  users'~(Alexa, 2018a). Alexa's data is imperfect, but is extensively  used in the web measurement literature.}. Upon downloading the homepage, we extracted the page meta description information (a short summary of the page's content provided by the site developer) and page title. Our population of pornography websites is comprised of sites with `porn' in the URL, meta description, or title of the page. `Porn' functions as an excellent identifier as the text is rarely used outside of the context of pornography as very few words other than `pornography' contain the letter sequence `porn.'

\subsection{Identifying Third-Parties on Websites}

To identify third-parties found on a given website we used the webXray software platform. webXray `is a tool for analyzing third-party content on web pages and identifying the companies which collect user data' (webXray, 2018). webXray functions by loading a given web page in the Chrome web browser. During the page load, webXray records all network traffic so that instances where user data is exposed to third-parties are identified. This network traffic is initially in a raw format and webXray `uses a custom library of domain ownership to chart the flow of data from a given third-party domain to a corporate owner, and if applicable, to parent companies' (webXray, 2018). For example, if a given page initiates a request to the domain `2mdn.net', webXray will reveal that the page hosts code from DoubleClick, a subsidiary of Google, which is in turn a subsidiary of Alphabet. webXray also records data on all cookies set in the browser during page loading. Overall,webXray provides ample data from which to investigate the nature and scope of tracking on popular pornography websites.

\subsection{Extracting and Analyzing Privacy Policies}

This study examines the role of consent in online tracking and we conducted an additional analysis of site privacy policies using policyXray, a companion program to webXray (Libert, 2018). Once a webXray analysis is completed, policyXray is used to locate the privacy policy of a given page by searching for links containing text such as `privacy' and `privacy policy'. policyXray then loads the privacy policy page in the Chrome web browser, injects the Mozilla Readability.js library into the page, and extracts the page's policy (Libert, 2018). The extracted policy is then analyzed to determine reading difficulty, time needed to read the policy, and if the third-parties detected collecting user data on the website are disclosed in the policy.

policyXray searches not only for the identified owner of a given tracker, but the parent companies as well, meaning the policy of a page which initiates a request to `2mdn.net' will be searched for `DoubleClick,' `Google,' and `Alphabet.' Likewise, policyXray accounts for spelling variations so that both `DoubleClick' (one word) and `Double Click' (two words) are searched. Overall, policyXray is designed to give as many chances as possible for disclosure to be counted and is intentionally generous in this regard (Libert, 2018).

\subsection{Content Analysis of Domain Names and `Sexual Interest'}

To determine the extent to which the domain names of sites in the sample could alone appear to reveal specific sexual/gender preference, identity, or sexual topic of interest of the site content or a site user, we conducted a content analysis of the site URLs. Content analysis is used for making valid and replicable inferences from texts to their context (Krippendorff, 1980). It is a useful method to employ when an individual investigator's reading of a text proves inadequate (Holsti,1969). We drew a representative random sample of 378 site URLs from the larger population of 22,484. Confidence Level for the sample was 95\% with a Confidence Interval of 5.

We used four coders from diverse backgrounds: one primary researcher and three volunteers. Three coders were women (one identified her sexuality as fluid; the others as queer), and one was a heterosexual man. Coders were trained using a code book with guidelines and examples for coding Presence or Absence of words or phrases that `reveal or strongly suggest to the average user' one or more specific gender/sexual identities or orientations, or topics of interest or focus. The `Presence'/`Absence' categories were defined a priori based on theoretical understandings of gender and sexuality. Coders were instructed to code \emph{Presence} for: `Any word or phrase that indicates or suggests the porn content will feature a specific gender or sexual identity, orientation, or preference\footnote{These might include proper, slang, and/or derogatory  words or phrases like: men, gay, heterosexual, lesbian, transgender,  dyke, chick.},' and/or `Any word or phrase that indicates or suggests the porn content will feature a specific sexual focus, body part or type, identity or character (like race, nationality, ethnicity,religion, profession), act, fetish, interest, porn genre, porn trope, etc.\footnote{These might include proper, slang, and/or derogatory words  or phrases like: feet, boobs, MILF, Latina, BBW, anal, incest, zoo,  rape, secretary.}' Coders were instructed to code \emph{Absence} indicating: `...the domain does NOT reveal or strongly suggest to the average user one or more specific gender or sexual: identities or orientations, and/or topic(s) of interest or focus. Instead, the domain indicates generic porn/adult themes\ldots{}\footnote{These might  include: style of porn (Amateur, VR, cartoon), xxx, adult, sex, hot,  mobile, chat, vids, tube, free.}' Definitions and examples were written to render masculinity and heterosexuality visible and thus not reinforced as normative\footnote{Coders were encouraged to not  categorize porn targeted to heterosexual men as generic or  \emph{Absence} (e.g. `girl' would be coded \emph{Presence}, as would  `boy;' `doggystyle' would be coded \emph{Presence}, as would  `bareback').}. During the 45-minute training, disagreements between coders were discussed until consensus was established; the code book was revised accordingly. All coders completed coding in less than one hour, minimizing concerns of coder fatigue. Krippendorff's alpha, a measure of reliability among coders, was .86, which falls within an acceptable range (Krippendorff, 2010).

\subsection{Limitations}

While we use a robust methodology, no study is without limitations. Regarding the construction of our list, while it is the largest number of pornography websites to be studied in the context of web tracking, it does not include all such websites, and due to the opaque nature of the Alexa list it is impossible to quantify how reliable the sample is overall. The Alexa list is used widely in the literature and thus our study inherits a common weakness. Regarding measuring third-parties with webXray, several limitations may apply.  First, due to a variety of factors including IP blacklisting and rate limiting, the computer running webXray may be identifying as a `bot' and blocked by some websites. Likewise, some types of third-party content may not load and will be missed by webXray. Overall, the measures produced by webXray should be taken as low-bound measures, as the actual amount of tracking may be higher. Regarding policyXray, limitations include the possibility extracted text does not correspond to the actual policy, portions of the policy may not be extracted, and policies may not load correctly due to issues related to being marked a `bot.' The content analysis has limitations typical of the method. Namely, Wimmer and Dominick (2011:159) note findings: ``are limited to the framework of the categories and the definitions used in that analysis. Different researchers may use varying definitions and category systems to measure a single concept''. We worked to account for our theoretical and political positionality in the defining of categories to make more transparent these researcher influences.

\section{Findings}

Our March 2018 analysis successfully examined 22,484 sites drawn from the Alexa list of one million most popular websites where the URL, page title, or page description includes `porn.' We found third-party tracking is widespread, privacy policies are difficult to understand and do not disclose such tracking, and third-parties may often be able to infer specific sexual interests based solely on a site URL.

\subsection{Third-Party Tracking}

Our results indicate tracking is endemic on pornography websites: 93\% of pages leak user data to a third-party; the pages that leak data do so to an average of seven domains; 79\% have a third-party cookie (often used for tracking); of the pages with cookies, there is an average of nine cookies; and only 17\% of sites are encrypted, allowing network adversaries to potentially intercept login and password details.\footnote{Note that even if a homepage does not use encryption,  a separate login page may; however, it is now common practice to  encrypt all pages.}

\begin{table}
\centering
\caption{Top Ten Third-Parties}
\begin{tabular}{llrr}
Company & \% Sites & Country & Porn-Focused \\
\hline
Google & 74 & United States & - \\
\textbf{exoClick} & 40 & Spain & Yes \\
Oracle & 24 & United States & - \\
\textbf{JuicyAds} & 11 & Netherlands & Yes \\
Facebook & 10 & United States & - \\
\textbf{EroAdvertising} & 9 & Netherlands & Yes \\
Cloudflare & 7 & United States & - \\
Yadro & 7 & Russia & - \\
New Relic & 6 & United States & - \\
Lotame & 6 & United States & - \\
\hline
\end{tabular}
\end{table}

We identified 230 different companies and services tracking users in our sample.  Such tracking is highly concentrated by a handful of major companies, some of which are pornography-specific. Of non-pornography-specific services, Google tracks 74\% of sites, Oracle 24\%, Facebook 10\%, Cloudflare and Yadro 7\%, and New Relic and Lotame 6\%. Porn-specific trackers in the top ten are exoClick (40\%), JuicyAds (11\%), and EroAdvertising (9\%). 171 companies and services are present on fewer than 1\% of sites, exhibiting a long-tail effect. Figure 1 illustrates data flows between five of the most popular porn websites and several third-parties.

\begin{figure*}
    \includegraphics[width=\textwidth]{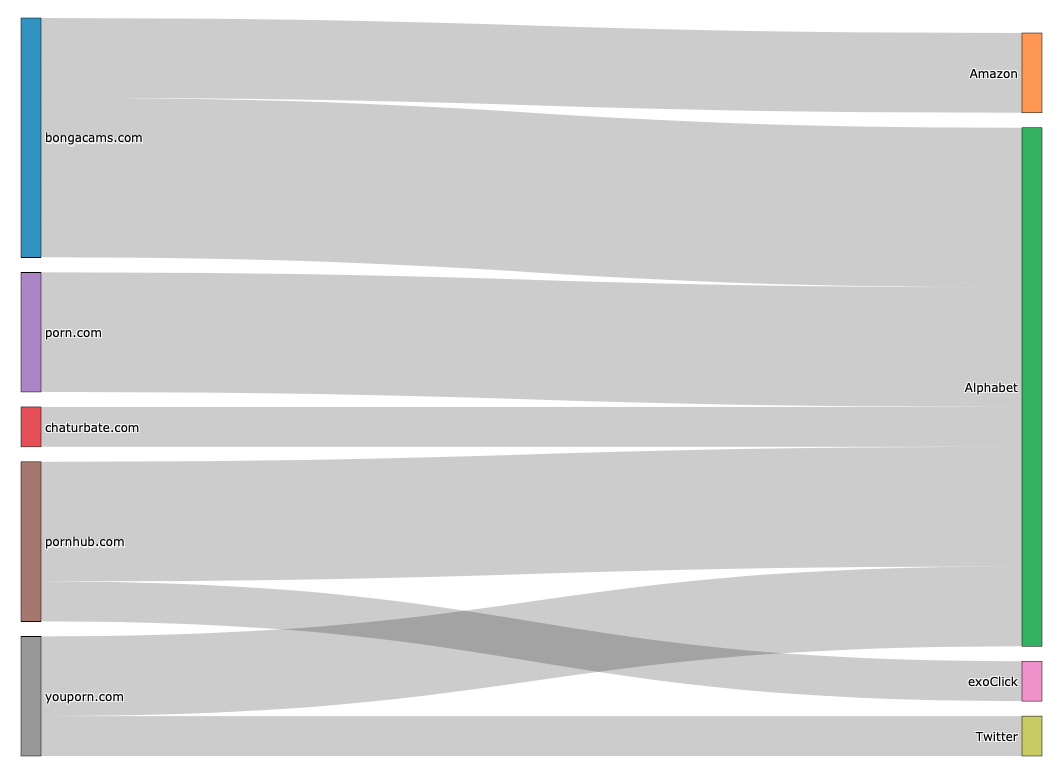}
    \caption{Diagram of data flows to third-parties on major porn sites.  Note Alphabet is the holding company of Google.}
\end{figure*}

\begin{table}
\centering
\caption{Breakdown of Google Services Used}
\begin{tabular}{ll}
Service Name & \% Sites \\
\hline
Google APIs & 50.1 \\
Google Analytics & 49 \\
DoubleClick & 11 \\
Google Tag Manager & 7 \\
Blogger & 2 \\
YouTube & 1 \\
AdSense & 1 \\
\hline
\end{tabular}
\end{table}

The majority of non-pornography companies in the top ten are based in the U.S., while the majority of pornography-specific companies are based in Europe. One reason may be differing cultural and commercial norms towards sexual content. In the U.S., many advertising and video hosting platforms forbid `adult' content. For example, Google's YouTube is the largest video host in the world, but does not allow pornography. However, Google has no policies forbidding websites from using their code hosting (Google APIs) or audience measurement tools (Google Analytics). Thus, Google refuses to host porn, but has no limits on observing the porn consumption of users, often without their knowledge. Table 2 is a breakdown of the use of Google services, and makes clear how Google's content policies have an impact on use for their services by pornography websites.

\subsection{Privacy Policies}

We successfully extracted privacy policies for 3,856 sites, 17\% of the total. Major reasons for not extracting the policy of a given site are that it does not have a privacy policy, the link for the policy uses uncommon phrasing, or the structure of the page makes it difficult to extract a policy URL (as with a modal window). We found policies are written at a grade level 14 on the Flesch-Kinkaid scale, meaning two years of college are estimated to be needed to understand the policy. Policies have an average word count of 1,750 and take seven minutes to read (McDonald and Cranor, 2008). Only 11\% of third-parties observed tracking users on a given page are listed in the policy, indicating users may have no means to learn which companies might have troves of data about their porn use. The difficulty of understanding a policy indicates those who do not have college-level education (and likely many that \emph{do}), may be unable to give informed consent on pornography websites. Additionally, if the names of companies collecting user data are missing, it is impossible for users to consent to the use of their data for tertiary purposes.

\subsection{Exposure of Sexual Interest}

Based on a random sample, 44.97\%\footnote{While 44.97\% is alarming,  the percentage may be even higher. Our 4 coders, although diverse,  could not possibly be aware of all sexual terms and slang in the URLs  analyzed. Likely some sites coded as generic actually contained  references undetected by coders without niche knowledge.} of porn site URLs expose or strongly suggest the site content includes or targets one or more specific gender or sexual: identities or orientations, and/or topic(s) of interest/focus. To elucidate: these porn domains contain words or phrases that would likely be generally understood as an indicator of a particular sexual preference or interest inherent in the site's content, these might also likely be assumed to be tied to the user accessing that content. As example, some sites coded reliably across coders as exposing such interests include: `http://bestialitylovers.com,' `http://boyfuckmomtube.com,' and `http://hdgayfuck.com.' The remaining sites in the sample do not make easily discernible the type of content on the site. Examples of these `generic' domains include `http://watch8x.com,' and `http://gohdporn.com.' While we reiterate specific types of porn do not necessarily indicate user gender/sexual identity or interest, these results reveal the extent to which third-parties might \emph{assume} users' specific sexual characteristics based on sites visited. Venturing further into a site would provide an even more complete understanding of the content therein.

\section{Discussion}

Below we present three primary implications drawn from the results. Each combines our findings with theoretical and empirical grounding to make an argument related to sexual data and online porn. First, we argue porn data leakage represents a unique and elevated risk compared to many other types of data. We base this argument on our quantitative results that reveal a large majority of our sample leaked users' sexual data to third-parties, combined with the growing precedent for high-profile, large-scale leaks, hacks, and missteps with sexual data. Next, we argue marginalized groups will likely be most targeted and harmed by such tracking. The extent to which gender and sexual interests could be inferred from site URLs demonstrates the troubling potential for the tracking and disciplining of sexual interests labeled non-normative. There is precedent for such targeted abuse of women and other marginalized populations online, and we contend their susceptibility to technological attacks based on moral outrage point to wider societal vulnerabilities in the face of constantly shifting socio-sexual norms. Finally, based on our privacy policy findings, we argue porn sites and other industrial actors dealing in this data must acknowledge they are engaged in a transaction involving sex and power, and thus require affirmative sexual consent from users.

\subsection{The Unique and Elevated Risks of Porn Data Leakage}

Most crucially, our results reveal the wide-scale privacy and security risks of consuming online pornography. The high percentage of site URLs that may reveal specific information about the content that users access constitutes an opportunity for the linking of this sensitive data to those users' other tracked online activities and profiles. Turow et al. (2015b) and Turow (2017) demonstrated how the retail industries' tracking and profiling of shoppers lead to discriminatory pricing practices and other privacy violations. Those studies argue for the severity of the risks of such tracking. We contend that the tracking of online porn consumption represents an even riskier violation of privacy, in line with Citron's (2019:1870, 1881) argument that:

\begin{quote}
Sexual privacy sits at the apex of privacy values because of its importance to sexual agency, intimacy, and equality. We are free only insofar as we can manage the boundaries around our bodies and intimate activities... It therefore deserves recognition and protection, in the same way that health privacy, financial privacy, communications privacy, children's privacy, educational privacy, and intellectual privacy do.
\end{quote}

\noindent While perhaps less publicly discussed than retail tracking, in recent years porn and other sites related to sexual data have grabbed headlines as they became targets of attack. In addition to the site URLs studied herein, porn websites store a plethora of customer information that, if the sites are hacked or if users are tracked online, could potentially reveal specific users' private sexual data. Between 2012 and 2018, at least 12 porn sites, sites tied to non-consensual voyeurism, and sites for extramarital affairs were visibly breached (Hunt, 2018). In some cases, these breaches caused significant harm to users and their families. The extra-marital site Ashley Madison suffered a prominent hack exposing 32 million names, credit card numbers, email and physical addresses as well as sexual interests of customers (Isidore and Goldman, 2015).

The particularly sensitive nature of data collected by porn websites, combined with their comparatively lax security, can prove irresistible to hackers. In a 2012 porn site attack, hackers said the site's mediocre security made it, `too enticing to resist' (Geuss, 2012). The hackers stole data of more than 73,000 subscribers including email addresses, passwords, usernames, and information from 40,000 credit cards (BBC,2012). Earlier that year, YouPorn suffered a major breach, exposing thousands of user emails and passwords that promptly began circulating online (BBC, 2012). A website related to more specific sexual practices, Rosebuttboard.com, a forum about `extreme anal dilation and anal fisting' was hacked in 2016, resulting in more than 100,000 user accounts being exposed (Cox, 2016). In 2018, thousands of users who accessed a bestiality website had personal details including email addresses, birth dates, and IP addresses circulated on public image boards by hackers (Cox, 2018).

Malicious advertisements, delivered by the same types of third-parties documented herein, have been found on many porn websites, including xhamster, Pornhub and xvideos (Lee, 2013). One malicious ad, running on Pornhub's website for over a year, could take control of a victim's computer and click on other fake ads to generate income for those operating the scheme (Griffin, 2017). Although malicious ads can be found on many websites, on porn sites they are less likely to be reported by users based on the nature of their visit to the site. Tracking across the web elevates the risks, especially considering the large corporate entities we discovered can potentially access user data through third-party requests, such as Google and Facebook. Our results, combined with the precedent for large-scale data leaks of online sexual data, illustrate the unique and elevated risks associated with porn data leakage and tracking.

\subsection{Targeting Difference: Tracking Sexual Data and Vulnerable Populations}

Our results indicate that the URLs of 44.97\% of the 22,484 sites in our sample likely reveal a specific gender/sexual preference, identity or interest related to the sites' content. We argue revelations about individuals' porn use and the specific sexual content of sites they visit are unlikely to publicly be granted the nuance due humans' complex sexual interests, curiosities and identities \footnote{Many domains in  our sample illustrate how quickly nuance might be lost in favor of  exposure, panic and severe consequences. Sites like `momboysex.ws,'  `freerapeporn.org,' and `pornwithanimals.net' would create scandal  amid revelations they were frequented by a public figure, as well as  personal/professional fallout for an ordinary citizen.}. This flattening is even more likely to occur online. As Marwick (2017: 180) noted regarding two prominent online sexual privacy controversies: `\ldots the same properties of social media that facilitate activism and cultural participation can \ldots enable networked abuse and targeted intimidation.' Those most likely to be impacted by online sexual privacy violations are traditionally marginalized and vulnerable communities, especially women, people of color, LGBTQ and other marginalized gender/sexuality communities, and those whose sexual interests are labeled deviant or abnormal in their community (Citron, 2019).

Sexual interests typically coded as (hetero)normative become invisible in the face of queer, `deviant,' or `alternative' interests marked as difference (Butler, 1993). That `difference' made visible against a constructed sexual `norm,' makes certain groups most vulnerable to moralistic attacks based on sexual data. Plummer (2003: 57) explains, `...the ``sexual citizen'' \ldots occupies a classed, ethnicized, gendered, and age-grouped position in society. That is, not all sexual citizens will be treated equally, fairly, in the same way\ldots' Franks' (2017: 431) notion of `intersectional surveillance,' building on Crenshaw's (1989) foundational work on intersectionality, emphasizes that `those subjected to multiple sources of subordination are also subjected to multiple sources of surveillance.' In what Banet-Weiser and Miltner (2016) call `networked misogyny,' women are singled out for a disproportionate share of online abuse. `Revenge porn,' or non-consensual pornography, in which nude or sexual photos or videos are spread without consent of the subject in order to intimidate and cause harm, has impacted at least 4\% of U.S. internet users, or 10.4 million people (Lenhart et al., 2016). Women and those who identify as lesbian, gay, or bisexual~are the most frequent victims of revenge porn and other forms of online harassment and sexual privacy violations (Lenhart et al., 2016; Citron, 2019). Chu and Friedland (2015: 3) explain that when women's privacy is violated and they are shamed online: `The message is... you deserve no protection, no privacy. ...Slut-shaming \ldots{}blames the user---her habits of leaking---for systemic (networked) vulnerabilities\ldots' Thus, surveillance and exposure are deployed to discipline difference.

Moral judgments can lead to devastating consequences for any sexual interests labeled `immoral' (Warner, 1999: 5), especially when these judgments are encoded into law. For example, same-sex relations between consenting adults are criminalized in 70 United Nations member states,with punishments ranging from imprisonment to death (Fox et al., 2019). The consequences of sexual privacy violations in such contexts would clearly be severe. Even in societies with less regulation around sex, breaches of sexual privacy often have bodily stakes. In the U.S., the Ashley Madison hackers were motivated by moral judgments about the site's intended user-base: people (mostly men) seeking extra-marital affairs. In the breach, hackers threatened to release users' names, addresses, credit card transactions, and `secret sexual fantasies' if the site was not shut down. When the hackers did publicly post the data, they claimed it was a public service exposing the `fraud, deceit, and stupidity' of users looking to cheat on their significant others (Isidore and Goldman, 2015). The data dump resulted in devastating outcomes for some users, including divorces, suicides, hate crimes, and extortion emails (Baraniuk, 2015; Segall, 2015). With primarily heterosexual men as victims, here we see that \emph{difference} coded broadly, and deemed immoral, widens the landscape of vulnerability. Considering popular perceptions of porn consumption as a masculine practice, combined with the frequent online abuse of women, one could imagine the immediate moral dimension to data breaches exposing women's sexual porn consumption. And while marginalized groups are most at risk, their vulnerabilities point to wider implications for a society in which ostensibly private sexual data is accessible to the extent defined in our results and combined with a growing precedent of technological attacks based on moral outrage and conditional understandings of the right to privacy.

\subsection{Sex, Porn, and the Complications of Consent}

We argue our findings reveal troubling violations by porn sites (and other industrial actors trading in porn consumption data), of users' sexual privacy and autonomy. Porn website privacy policies are long, dense, difficult to understand, and only 11\% of the third-parties observed tracking users on a given page are listed in the policy, leaving users ignorant of which organizations may be assembling catalogues of their perceived sexual interests. The potentially catastrophic and certainly often violent consequences of past (and undoubtedly future) sexual privacy breaches are reflective of problematic norms of sexual consent in a patriarchal and heteronormative rape culture. The language of `consent' is often mobilized in online privacy policies and other legalese related to the use of personal data. Yet, our results point to parallels in the dubious understandings of consent seen in intimate sexual encounters \emph{and} in online sexual activity. We argue both in-person and online sexual activity alike (including online porn consumption) must be conducted only when\emph{ affirmative consent} has been granted. In the case of private sexual data, such as that which leaks from porn websites, the limitations of the user's ability to consent are currently so egregious they create numerous opportunities for violence, blackmail, and discrimination, and constitute a clear and present danger to those who consume online pornography.

In interpersonal interactions, sexual consent is often mistaken to mean `no means no, and silence means yes.' However, policy makers are increasingly working to (re)define sexual consent as necessarily\emph{affirmative}: silence does not equal consent, someone must\emph{ communicate} their consent for true sexual autonomy. Further, Citron (2019: 1882) notes, `Consent facilitated by sexual privacy is contextual and nuanced - it does not operate like an on-off switch.' In 2015, all New York schools adopted an affirmative definition of sexual consent, stating: `Consent can be given by words or actions, as long as those words or actions create clear permission regarding willingness to engage in the sexual activity. Silence or lack of resistance, in and of itself, does not demonstrate consent' (Delamater, 2015: 591). Sexual consent is increasingly, and we believe usefully, conceptualized in this way.

In most U.S. online privacy policies consent is ideally, `based on the idea that individual social media users make conscious, rational, and autonomous choices about the disclosure of their personal data' (Custers et al., 2014: 270). However, the actual use of personal data usually does not live up to this ideal. Often websites (that somewhere contain a privacy policy) assume user consent to any potential outcomes of their use of the site, despite users not \emph{saying} or \emph{doing} anything signaling agreement. And indeed, most users interact very little or not at all with privacy policies and tools for consenting to the use of personal data (Larose and Rifon, 2006; Custers et al., 2014). 52\% of Americans do not understand what a privacy policy \emph{is} (Smith, 2014). A nationwide survey found 65\% of respondents mistakenly believed if a site contains a privacy policy, their information will not be shared without permission (Turow et al., 2015a). In fact, there is no legal requirement in the U.S. for a website to have a privacy policy, and the mere existence of a policy does not mean user privacy is\emph{ protected}, only that users are \emph{informed} of practices. In the U.S. it is perfectly legal to have a privacy policy explicitly stating porn browsing data will be sold. Thus, consent online is at best opaque and misunderstood, and at worst, intentionally deceptive.

Sexual consent is carried out within systems of power, and consent becomes especially murky when there is a power imbalance between participants. Dougherty (2015: 238) argues consent is like a promise, and promises can `protect individuals from imbalances of power within a relationship.' Our results reveal a troubling power imbalance in the negotiation of users' private sexual data. Some of the world's largest and most powerful corporations have access to this data. Not only are users often unable to truly give consent to the collection and use of their private sexual data; if an `agreement' \emph{is} breached, the power difference between the user and corporation is so vast the user has little recourse or protection.

The similarities between online and offline violations of consent have not gone unnoticed online where metaphors of sexual violence proliferate. Chun and Friendland (2015: 15), citing internet culture examples like `frape,' `\#rapeface,' and `pwned,' argue the word `rape' is increasingly popular online `\ldots to suggest the possibility of online subjects dominating, violating, or transforming other online subjects.' The consequences of exposure of sexual data certainly imply the violence associated with such internet parlance. In the case of `sextortion,' in which people are blackmailed into performing sexual acts so attackers won't release their sexual images/data, Citron (2019:1925) notes, `victims have described feeling like they were ``virtually raped.'''

Alarmingly, regulators often decline to act in the face of user vulnerability to privacy violations:

\begin{quote}
With the exception of certain forms of health and financial information\ldots{} companies continue to be free to collect huge streams of individuals' data\ldots{} to construct profiles of individuals with the data, to share the data, and to treat people differently\ldots{} based on conclusions drawn from those hidden surveillance activities. (Turow et al., 2015a: 8)
\end{quote}

\noindent The reality of user vulnerabilities when consuming porn online, especially considering the leakage and data specificity identified herein, combined with our privacy policy results, lead us to argue that the ability to consent to the collection, use, or exposure of this personal sexual data constitutes a form of sexual consent. Thus, it ought to be as carefully considered, defined, and regulated as interpersonal sexual consent\footnote{Interpersonal sexual consent,  though more attended to than the issues we highlight here, also  remains poorly defined and regulated (Citron, 2019).}. Dougherty (2015: 227) argued affirmative consent is needed, `in the case of high-stakes consent...with sexual consent as a paradigm of high-stakes consent.' We can not imagine many online privacy concerns more high-stakes than those identified in this study, and so we argue users' affirmative consent should be obtained by porn sites holding such sensitive data\footnote{How such consent might be conceptualized, and  ultimately, implemented, is beyond the scope of this article; we  believe recognition of the stakes we've described can be a useful  starting point.}. As in any sexual interaction, silence must not be mistaken for consent, and individuals should have a clear understanding of the power dynamics of the sexual exchange they are entering when visiting porn sites, as well as the procedures for withdrawing consent.\footnote{See Custers (2016) on the need for expiration dates  for online informed consent.}

\section{Conclusion}

RQ1 asked to what extent porn websites reveal user data and allow for third-party tracking. We have demonstrated they leak data through third-party requests to a large and concerning extent. RQ2 asked who has access to the data and whether the access is disclosed in the sites' privacy policies. We have shown the data are often accessed by large corporations, parties usually \emph{not} identified in porn site privacy policies as having access to user data. RQ3 asked whether users' sexual interests might be revealed in the surveillance and tracking of porn usage. The user data often suggests or reveals gender/sexual identities or interests represented in the porn site URL accessed, and thus poses an additional risk if tracked and assumptions about users' sexual identities/interests are linked to personal identifying information. RQ4 asked about the potential implications of porn website tracking and surveillance. Through our results and connections to past porn site privacy and security breaches and controversies, we demonstrate that the singularity of porn data and the characteristics of typical porn websites' lax security measures mean this leakiness poses a unique and elevated threat. We have argued \emph{everyone} is at risk when such data is accessible without users' consent, and thus can potentially be leveraged against them by malicious agents acting on moralistic claims of normative gender or sexuality. These risks are heightened for vulnerable populations whose porn usage might be classified as non-normative or contrary to their public life. Finally, we argue porn sites currently operate with an unethical definition of sexual consent considering the sensitive sexual data they hold. We contend the overwhelming leakiness and sexual exposure revealed in our results mean porn sites ought to better account for user security as well as adopt policies based on affirmative consent.

Despite these risks, maintaining private access to online porn is crucial for all. Porn often serves educational, exploratory and liberatory -- or `sex positive' -- outcomes. When sexual `norms' are potentially upheld and leveraged by the state or other agents, every citizen is ultimately at risk of being labeled `non-normative' or pathological. Such online privacy risks and offline bodily and reputational risks can lead to unequal power relations and the moving underground -and associated marginalization- of minority (a constantly in-flux and culturally-contingent category) sexual interests and identities. Others have theorized approaches to online and sexual privacy concerns that we find useful and promising in light of our results\footnote{See Citron (2019) on sexual privacy, Plummer's (2003)  concept of `intimate citizenship,' and Baruh and Popescu's (2017: 592)  argument that regulatory efforts ought to `recognize the nature of  privacy as both a collective value and a collective social  phenomenon,' rather than traditional individualist understandings of  privacy self-management.}. And we recognize the theoretical and political power of Chun and Friedland's (2015: 3) contention that:

\begin{quote}
We need to fight for the right to take risks---to be in public---and not be attacked. \ldots(and) by building new forms of interaction that cannot `leak' because they do not seek to create imaginary bubbles of privacy between users in the first place.
\end{quote}

\noindent However, our results reveal just how susceptible our sexual data is to accidental \emph{or} hostile exposure and thus, to moral targeting, on- and offline violence, and other negative consequences. Our data lend weight to Citron's (2019: 1877) warning that, `Thanks to networked technologies, sexual privacy can be invaded at scale and from across the globe.' While we should always work toward exposing the constructed and complex nature of `normalcy' (Warner, 1999), we have demonstrated an imperative to swiftly and pragmatically address and disrupt the current and ongoing widespread leakiness of online porn sites.

While the findings of this study are far from encouraging, we do believe regulatory intervention may have positive outcomes. The form of consent currently found in U.S. self-regulatory `opt-out' systems fails to meet sexual consent norms and reinforces the `blame the victim' mentality that often emerges in slut shaming and other forms of sexual violence.  Scholars are not immune from such attitudes as evidenced by the `privacy paradox' which implicitly holds users responsible for privacy violations rather than the powerful actors conducting surveillance. In contrast, the European Union's GDPR formulation of online tracking consent more closely matches norms for sexual consent by emphasizing consent must be affirmative \emph{and} freely given, thereby providing greater protections to users. Our results demonstrate the imperative to attend to outcomes of the GDPR and to develop models of affirmative digital consent for porn websites that meet the diverse requirements for providing and withdrawing consent in sexual interactions.

\section{References}

Alexa (2018a) How are Alexa's traffic rankings determined? Available at: https://support.alexa.com/hc/en-us/articles/200449744-\%20How-are-Alexa-s-traffic-rankings-determined- \\

Alexa (2018b)~\emph{The top 500 sites on the web}.~Available at: https://www.alexa.com/topsites \\

Banet-Weiser S and Miltner KM (2016) \#MasculinitySoFragile: culture, structure, and networked misogyny. \emph{Feminist Media Studies} 16(1): 171-174. \\

Baraniuk C (2015) Ashley Madison: `suicides' over website hack. \emph{BBC News,} 24 August. Available at: https://www.bbc.com/news/technology-34044506 \\

Barker M (2014) The `problem' of sexual fantasies. \emph{Porn Studies} 1(1-2): 143-160. \\

Barth S and de Jong MDT (2017) The privacy paradox -- investigating discrepancies between expressed privacy concerns and actual online behavior -- a systematic literature review. \emph{Telematics and Informatics, 34}: 1038-1058. \\

Baruh L and Popescu M (2017) Big data analytics and the limits of privacy self-management. \emph{New Media \& Society 19}(4): 579-596. \\

BBC News (2012) Porn site breached in hack attack. Available at: https://www.bbc.com/news/technology-17339508 \\

Binns R, Lyngs U, Van Kleek M, et al. (2018) Third party tracking in the mobile ecosystem. In: \emph{WebSci '18: 10th ACM Conference on Web Science}, Amsterdam. \\

Butler J (1993) \emph{Bodies that Matter}. New York: Routledge. \\

Campbell JE and Carlson M (2002) Panopticon.com: online surveillance and the commodification of privacy. \emph{Journal of Broadcasting \& Electronic Media} 46(4): 586-606. \\

Chun W and Friedland S (2015) Habits of leaking: of sluts and network cards. \emph{Differences 26}(2): 1--28. \\

Citron DK (2019) Sexual privacy. \emph{The Yale Law Journal} 128: 1870-1960. \\

Cox J (2016) Another day, another hack: is your fisting site updating its forum software? \emph{Motherboard, 10 May}. Available at: https://motherboard.vice.com/en\_us/article/qkjj4p/rosebuttboard-ip-board \\

Cox J (2018) Thousands of bestiality website users exposed in hack. \emph{Motherboard, 29 March}. Available at: https://motherboard.vice.com/en\_us/article/evqvpz/bestiality-website-hacked-troy-hunt-have-i-been-pwned \\

Cranor LF, Reagle J and Ackerman MS (2000) Beyond concern: understanding net users' attitudes about online privacy. \emph{The Internet upheaval: raising questions, seeking answers in communications policy, pp.47-70}. \\

Crenshaw K (1989) Demarginalizing the intersection of race and sex: a Black feminist critique of antidiscrimination doctrine, feminist theory and antiracist politics. \emph{University of Chicago Legal Forum}: 1(8): 139-167. \\

Custers B (2016) Click here to consent forever: expiry dates for informed consent. \emph{Big Data \& Society 3}(1). \\

Custers B, van der Hof S and Schermer B (2014) Privacy expectations of social media users: the role of informed consent in privacy policies. \emph{Policy \& Internet 6}(3): 258-295. \\

Delamater C (2015) What ``yes means yes'' means for New York schools: the positive effects of New York's efforts to combat campus sexual assault through affirmative consent. \emph{Albany Law Review 79.2}: 591. \\

Dougherty T (2015) Yes means yes: consent as communication. \emph{Philosophy \& Public Affairs} 43(3): 224-253. \\

Englehardt S and Narayanan A (2016) Online tracking: a 1-million-site measurement and analysis. \emph{Proceedings of the 2016 ACM SIGSAC Conference on Computer and Communications Security (pp. 1388-1401).} \\

Felten EW and Schneider MA (2000) Timing attacks on web privacy. \emph{In Proceedings of the 7th ACM conference on Computer and Communications Security (pp. 25-32)}. \\

Fox K, Pettersson H and Mackintosh E (2019) Where being gay is illegal around the world. \emph{CNN}, 8 April. \\

Franks MA (2017) Democratic surveillance. \emph{Harvard Journal of Law \& Technology }\emph{30(2): 425-489.} \\

Geuss M (2012) Porn site Digital Playground hacked, hackers say ``too enticing to resist.'' \emph{Ars Technica}. Available at: https://arstechnica.com/information-technology/2012/03/porn-site-digital-playground-hacked-hackers-say-too-enticing-to-resist/ \\

Greene D and Shilton K (2018) Platform privacies: governance, collaboration, and the different meanings of ``privacy'' in iOS and Android development. \emph {New Media \& Society} 20(4): 1640-1657. \\

Griffin A (2017) Pornhub hack: millions of people using adult video site could have been spied on. \emph{The Independent, 11 October.} \\

Gross L (2001) \emph{Up from Invisibility: Lesbians, Gay Men, and the Media in America.} New York: Columbia University Press. \\

Habib H, Colnago J, Gopalakrishnan V, et al. (2018) Away from prying eyes:analyzing usage and understanding of private browsing. \emph{Proceedings of the Fourteenth Symposium on Usable Privacy and Security}, pp.159-175. \\

Holsti OR. (1969) Content Analysis for the Social Sciences and Humanities. Reading: Addison-Wesley. \\

Hunt T (2018) Pwned websites. Available at: https://haveibeenpwned.com/PwnedWebsites \\

Isidore C and Goldman G (2015) Ashley Madison hackers post millions of customer names. \emph{CNN Tech}, 18 August. \\

Kennedy H (2016) \emph{Post, Mine Repeat: Social Media Data Mining Becomes Ordinary.} London: Palgrave \\

Kleinman A (2017) Porn sites get more visitors each month than Netflix, Amazon and Twitter combined. \emph{Huffington Post}, 4 May. Available at: https://www.huffingtonpost.com/2013/05/03/internet-porn-stats\_n\_3187682.html \\

Krippendorff K (1980) \emph{Content Analysis: An Introduction to its Methodology}. London:Sage. \\

Krippendorff K (2010) Krippendorff's alpha. In: Salkind, NJ (ed) \emph{Encyclopedia of}\emph{Research Design.} Thousand Oaks: Sage, pp.669-673. \\

Krishnamurthy B and Wills CE (2006) Generating a privacy footprint on the Internet. \emph{In Proceedings of the 6th ACM SIGCOMM Conference on Internet Measurement (pp. 65-70).} \\

LaRose R and Rifon N (2006) Your privacy is assured -- of being disturbed: websites with and without privacy seals. \emph{New Media \& Society} 8(6): 1009-1029. \\

Lee D (2013) Top porn sites `pose growing malware risk' to users. \emph{BBC News}, 12 April. Available at: https://web.archive.org/web/20141125065546/http://www.bbc.co.uk/news/technology-22093141 \\

Lenhart A, Ybarra M and Price-Feeney M (2016) \emph{Nonconsensual image sharing: one in}\emph{ 25 Americans has been a victim of ``Revenge Porn.''} Data \& Society Research Institute. Available at: https://datasociety.net/pubs/oh/Nonconsensual\_Image\_Sharing\_2016.pdf  \\

Libert T (2015) Exposing the hidden web: an analysis of third-party HTTP requests on one million websites. \emph{International Journal of Communication} 9: 3544-3561. \\

Libert T (2018) An automated approach to auditing disclosure of third-party data collection in website privacy policies. \emph{In Proceedings of the 2018 World}\emph{Wide Web Conference on World Wide Web} (pp. 207-216). \\

Libert T, Nielsen RK (2018) Third-Party web content on EU news sites: potential challenges and paths to privacy improvement. \emph{Reuters Institute for the Study of}\emph{Journalism}, University of Oxford. \\

Marwick AE (2017) Scandal or crime? Gendered privacy and the celebrity nude photo leaks. \emph{Ethics and Information Technology} 19(3): 177-191. \\

McDonald AM and Cranor LF (2008) The cost of reading privacy policies. \emph{I/S: A}\emph{Journal of Law and Policy for the Information Society, 4, p. 543.} \\

Mowlabocus S (2010) Porn 2.0? Technology, social practice, and the new online porn industry. In: Attwood F (ed) \emph{Porn.com: Making Sense of Online Pornography}. New York: Peter Lang, pp. 90-113. \\

Nissenbaum H (2010) \emph{Privacy in Context: Technology, Policy, and the Integrity of Social }\emph{Life}. Stanford: Stanford University Press. \\

Paasonen S (2018) \emph{Many Splendored Things: Thinking Sex and Play}. London: Goldsmiths Press. \\

Plummer K (2003) \emph{Intimate Citizenship: Private Decisions and Public Dialogues}. Seattle: University of Washington Press. \\

Pornhub (2017) Celebrating 10 years of porn\ldots and data! Available at https://www.pornhub.com/insights/10-years \\

Pornhub (2018) \emph{2017 year in review}. Available at:  https://www.pornhub.com/insights/2017-year-in-review \\

Ruckenstein M and Pantzer M (2017) Beyond the Quantified Self: thematic exploration of a dataistic paradigm. \emph{New Media \& Society} 19(3):  401-418. \\

Segall L (2015) Pastor outed on Ashley Madison commits suicide. \emph{CNN}\emph{Tech}, 8 September. \\

Sloop JM (2004) \emph{Disciplining Gender: Rhetorics of Sex Identity in Contemporary U.S. }\emph{Culture}. Amherst: University of Massachusetts Press. \\

Smith A (2014) Half of online Americans don't know what a privacy policy is. \emph{Pew}\emph{Research Center}. Available at: http://www.pewresearch.org/fact-tank/2014/12/04/half-of-americans-dont-know-what-a-privacy-policy-is/ \\

Smith C and Attwood F (2014) Anti/pro/critical porn studies. \emph{Porn Studies 1-2}: 7-23. \\

Tiidenberg K and Paasonen S (2018) Littles: affects and aesthetics in sexual age-play. \emph{Sexuality \& Culture} 23(2): 375-393. \\

Turow J (2012) \emph{The Daily You: How the New Advertising Industry is Defining Your}\emph{ Identity and Your Worth}. New Haven: Yale. \\

Turow J (2017) \emph{The Aisles Have Eyes: How Retailers Track Your Shopping, Strip Your}\emph{ Privacy, and Define Your Power.} New Haven: Yale. \\

Turow J, Hennessy M and Draper N (2015a) The trade off fallacy: how marketers are misrepresenting American consumers and opening them up to exploitation. Available at: https://www.asc.upenn.edu/sites/default/files/TradeoffFallacy\_1.pdf \\

Turow J, McGuigan L and Maris ER (2015b) Making data mining a natural part of life: physical retailing, customer surveillance and the 21st century social imaginary.\emph{ European Journal of Cultural Studies} 18(4-5): 464-478. \\

van Dijck J (2014) Datafication, dataism and dataveillance: Big Data between scientific paradigm and ideology. \emph{Surveillance \& Society} 12(2): 197-208 \\

Warner M (1999) \emph{The Trouble with Normal: Sex, Politics, and the Ethics of Queer Life}. Cambridge: Harvard University Press. \\

webXray (2018) webXray v 2.0 GNU General Public License v3.0 https://webxray.org \\

Williams L (ed) (2004) \emph{Porn Studies}. Durham: Duke University Press. \\

Wimmer RD and Dominick JR (2011) \emph{Mass Media Research: An Introduction} (9th ed). Boston: Wadsworth. \\

\end{document}